\documentclass[12pt]{iopart}

\usepackage{graphicx} 

\begin{document}

\title[Relation between Boltzmann and Gibbs entropy and example with multinomial dist.]{Relation between Boltzmann and Gibbs entropy and example with multinomial distribution}

\author{Pa\v{s}ko \v{Z}upanovi\'{c}}

\address{Faculty of Science, University of Split, R. Bo\v{s}kovi\'{c}a 33, 21000 Split, Croatia}
\ead{pasko@pmfst.hr}

\author{Domagoj Kui\'{c}}

\address{Faculty of Science, University of Split, R. Bo\v{s}kovi\'{c}a 33, 21000 Split, Croatia}
\ead{dkuic@pmfst.hr}

\begin{abstract}
  	General relationship between mean Boltzmann entropy and Gibbs   entropy is established. It is found that their difference is equal to fluctuation entropy, which is a Gibbs-like entropy of macroscopic quantities.   The ratio of the fluctuation entropy and mean Boltzmann, or Gibbs entropy  vanishes in the thermodynamic limit for a system of distinguishable and independent particles. It is argued that large fluctuation entropy  clearly indicates the limit where standard statistical approach should be modified, or extended using other methods like renormalization group. 
\end{abstract}

\section{Introduction}
\label{Introduction}

Entropy is the fundamental physical quantity in thermodynamics and statistical physics.  It was introduced by Clausius \cite{Clausius} in thermodynamics. It is a function of the macroscopic state of the system.   In contrast to   the Kelvin and the Clausius formulations of the second law in terms of processes, the second law expressed by entropy is formulated using a function of state. Boltzmann gave a microscopic definition of entropy, as a logarithm of the number of microscopic states that share the values of physical quantities of the macroscopic state of the system. 
Later, Gibbs gave another definition of entropy via probabilities of microscopic states of the system. 
Both definitions are used in some textbooks on statistical physics \cite{Landau},\cite{Kittel}. In some textbooks, we find only Boltzmann \cite{Reif} or Gibbs \cite{Feynman} definition of entropy. In these textbooks there is, either, no connection between these two definitions, or it is not established in a completely transparent way \cite{Landau}.

The problem of two different definitions of entropy is the subject of several works.  The reference \cite{Jaynes} puts into relation the Gibbs and Boltzmann $H$ functions. These are given respectively by
	\begin{eqnarray}
	H_G=\int f_N \ln f_N  d^3x_1 d^3p_1 \cdots d^3x_ N d^3p_N,\\
	H_B=N \int f_1 \ln f_1d^3x_1 d^3p_1 . 
	\end{eqnarray}
Here,  $f_N({\bf x}_1,{\bf p}_1, \dots , {\bf x}_N,{\bf p}_N; t)$ is the $N$-particle  probability density function in $6N$-dimensional phase space, while  $f_1({\bf x}_1,{\bf p}_1 ; t)$ is the single particle probability density function in $6$-dimensional phase space of one particle. They  are connected by the relationship \cite{Jaynes}:
\begin{equation}
f_1 ({\bf x}_1,{\bf p}_1; t) = \int  f_N({\bf x}_1,{\bf p}_1, \dots , {\bf x}_N, {\bf p}_N; t )  d^3x_2 d^3p_2 \cdots d^3x_ N d^3p_N .
\end{equation}
Jaynes has proven in \cite{Jaynes} 	that  $H_G \geq H_B$, where the difference between $H_G$ and $H_B$ is not negligible for any system in which interparticle forces have any observable effect on the thermodynamic properties. Furthermore, based on the use of the Gibbs canonical ensemble, in \cite{Jaynes} it is argued that the Gibbs entropy $S_G = -kH_G$, and not Boltzmann's $S_B = -kH_B$, is the correct statistical mechanics form that agrees with thermodynamic entropy. On the other hand, reference \cite{Jaynes} also makes a connection between the Gibbs entropy $S_G$ and the other definition of entropy which is also known as Boltzmann's, as a logarithm of the number of microscopic states that correspond to the macroscopic state of the system $S_B = k\ln W$, where $k$ is the Boltzmann constant. In this paper, we consider only  these two definitions. For the sake of simplicity, we call them Gibbs and Boltzmann entropy, respectively.

For clarity, it is also important to mention the reference \cite{Miranda}, where a numerical calculation of the Boltzmann entropy and the so-called Gibbs volume entropy, as candidates for the microcanonical entropy, is done on small clusters for two level systems and  harmonic oscilators. However, in view of recent discussions \cite{DH, FW, SW, HHD, Campisi} on the question which of these two definitions of  microcanonical entropy is justified on thermodynamic basis, numerical results of reference \cite{Miranda} are not decisive. The Gibbs volume entropy, and the Gibbs entropy defined in the sense that is used in our paper, coincide only if the microstates of energy less than or equal to the value $E$ are only allowed and they have equal probabilities.

The reference \cite{Attard} is interesting because it introduces an assumption of some internal structure of microstates. The author P. Attard in this way defines the entropy of a microscopic state, which is basically the logarithm of the weight (weight is proportional to but not equal to probability) assigned to the microscopic state. In the next step, the entropy of the macroscopic state is defined in an analogous way. Than the author claims that the entropy of statistical mechanics is the logarithm of the sum of all weights, which is equal regardless if one sums the weights over all microscopic, or over all macroscopic states. In this way,  this (so called total) entropy differs from the Gibbs-Shannon entropy by an additional term that is equal to the average entropy of microscopic states. The first problem with this approach are the entropies of microscopic states.  Disregarding that this is a departure from the standard approach that assumes zero entropy for a microscopic state, the main problem with it is that the weight of the microscopic state is not a uniquely defined quantity such as the probability, but only proportional to it. The second problem concerns with the definition of entropy as the logarithm of the total weight of all macroscopic states. In that way, one can not assign this entropy to any particular macroscopic state. In contrast to our analysis, which shows that if the Boltzmann and Gibbs entropies are defined in the sense used throughout this paper then they lead practically to same results in most cases, reference \cite{Attard} denies the validity of the use of Gibbs-Shannon form of entropy in statistical physics. 

Attard is not the only author who has introduced the entropy of a microscopic state. This was done earlier  by Crooks 
\cite{Crooks} who defined, based on information theoretic arguments, the entropy of microscopic state of a system 
as $-\ln p(x)$,  where $x$  specifies the positions  and momenta of all the particles. According to Crooks, this is the amount of information required to describe the microscopic state given that its occurs with a probability $p(x)$.  Crooks definition of entropy of microscopic state is based on the state probability which is a uniquely defined quantity. On the other hand,  Attard's definition \cite{Attard} relies on the ambiguously defined weight of a microscopic state. The definition given by Crooks enabled him to derive a general version of the fluctuation theorem for stochastic, microsocopically reversible dynamics \cite{Crooks}. 

Furthermore, Seifert \cite{Seifert} has adapted Crooks definition to the stochastic trajectory of a colloidal particle, using the Fokker-Planck equation that along with the initial conditions determines the probability as a function of the space and time coordinates of a particle.  In this way Seifert \cite{Seifert} has derived fluctuation theorem for arbitrary initial conditions and arbitrary time dependent driving over a finite time interval.


 In information theory, Shannon had searched for the function that has to describe information on the set of random events whose probabilities are $(p_1, \dots , p_n )$, under well known consistency conditions \cite{Shannon, JaynesBook}.
 He proved that the only function that satisfies the imposed conditions is, up to a multiplication factor,   
 \begin{equation}
 H_n =-\sum_{i=1}^{n}p_i \log p_i.
 \end{equation}
 It was named information entropy. It is equal to the Gibbs definition of entropy (\ref{eqgibbsentgen}).
 It is interesting to note that two completely different approaches  lead to the same expression. 
More about Gibbs-Shannon definition of   entropy interested reader can find in references \cite{JaynesBook, Caticha}. We note here only that Jaynes took the 
identity between these definitions as the starting point for his principle of maximum entropy known as MaxEnt, and then using this principle derived the canonical and grand canonical distributions \cite{Jaynes57}, first introduced by Gibbs \cite{Gibbs}. As an important example of a successful application of this principle in standard statistical physics, we refer to the paper written by Lee  and Presse \cite{Lee}, where they show that MaxEnt formulation for a small subsystem that exchanges energy with the heat  bath, follows from the partial maximization of the Gibbs-Shannon entropy of the total, isolated large system, over the bath degrees of freedom.

 The aim of this paper is to establish the relationship  between the Boltzmann and Gibbs definitions of entropy in a clear way, without putting in question these standard definitions.

The paper is organized in the following way.
In the next section general relationship between Boltzmann and Gibbs definition of entropy is established. By doing this, the condition for applicability of standard statistical description is derived. 

For pedagogical reasons it is shown, in the third section, that Boltzmann and Gibbs definition of entropy lead to the same result for a system with $N$ distinguishable and independent particles with two equally probable single particle states. 

The result of the third section is generalized in fourth section to the system of $N$ distinguishable and independent particles with $m$ single particle states with nonuniform  probabilities. 

Finally, in the  last section the results are summarized.

\section{General relationship between Boltzmann and Gibbs entropy}
\label{General}
We consider a general macroscopic system which can be in different macroscopic states that form together a set $A$. In order to be more specific, each macroscopic state $a \in A$ is determined by the values of $n$ macroscopic quantities $(a_1, \dots , a_n)$ defining it.   Let us assume that for any macroscopic state $a \in A$ the system of interest can be in $W(a)$ possible configurations (the multiplicity) which are denoted by $i_a = 1, \dots, W(a)$. Also, all configurations corresponding to a macroscopic state $a$ have equal probabilities $p_{i_a} = p(a)$. This is the principle of equal a priori probabilities applied to configurations corresponding to macroscopic state $a$. Then the probability of a macroscopic state $a$ is equal to
\begin{equation}
P(a) = W(a) p(a) . \label{eqmacroprobmultconfigprobgen}
\end{equation}
According to the Boltzmann definition, entropy of a system in a macroscopic state $a$, apart from the multiplicative factor given by Boltzmann constant $k$, is equal to 
\begin{equation}
S_B = \ln W(a) . \label{eqboltentgen}
\end{equation}

Entropy, like other thermodynamic quantities, in equilibrium state assumes its mean value. This holds in a microcanonical ensemble, too. Clearly, formulation of the second law for isolated systems based on Clausius inequality relies on the assumption that the relative difference between the most probable, which is at the same time the maximum, value of entropy, and a mean value of the entropy vanishes in the thermodynamic limit, $N \rightarrow \infty $ and $N/V = const.$ The reason is that, in the thermodynamic limit, the mean values of macroscopic quantities correspond closely to their values associated to the most probable macroscopic state. In the microcanonical ensemble, the most probable macroscopic state can be realized by overwhelmingly greatest number $W_{\mathrm{max}}$ of microscopic configurations of the system, and therefore, in that state the Boltzmann entropy $\ln W_{\mathrm{max}}$ is maximum. In other ensembles, nonuniform probabilities of individual configurations also come into play. One of the results of this paper shows that, in the thermodynamic limit, the relative difference between the most probable value and the mean value of Boltzmann entropy vanishes for all systems whose probability (\ref{eqmacroprobmultconfigprobgen}) of a macroscopic state is given by the multinomial distribution. 

The mean value of the Boltzmann entropy is obtained by averaging $S_B$ with respect to the probability of macroscopic states $P(a)$:
\begin{equation}
\overline{S}_B = \sum_{a \in A} P(a) \ln W(a) , \label{eqmeanboltentgen}
\end{equation}
where the summation goes over all macroscopic states $a \in A$. Using (\ref{eqmacroprobmultconfigprobgen}) we can write $\overline{S}_B$ as
\begin{equation}
\overline{S}_B = - \sum_{a \in A} W(a) p(a) \ln p(a) + \sum_{a \in A} P(a) \ln P(a) . \label{eqmeanboltentgen1}
\end{equation}
Remembering that all configurations belonging to a macroscopic state $a$ have equal probabilities $p(a) = p_{i_a}$, for $i_a = 1, \dots, W(a)$, we then have from (\ref{eqmeanboltentgen1}) that
\begin{equation}
\overline{S}_B = - \sum_{a \in A}\sum_{i_a = 1}^{W(a)} p_{i_a} \ln p_{i_a} + \sum_{a \in A} P(a) \ln P(a) . \label{eqmeanboltentgen2}
\end{equation}
The first term in (\ref{eqmeanboltentgen2}) is the Gibbs entropy of the system whose probability distribution of configurations is $p_{i_a}$, $a \in A, i_a = 1, \dots, W(a)$,
\begin{equation}
S_G = - \sum_{a \in A}\sum_{i_a = 1}^{W(a)} p_{i_a} \ln p_{i_a} . \label{eqgibbsentgen}
\end{equation}
The summation in (\ref{eqgibbsentgen}) goes over all configurations of the system. From (\ref{eqmeanboltentgen2}) and (\ref{eqgibbsentgen}) we then obtain a general relationship between the Gibbs entropy $S_G$ and the mean value of the Boltzmann entropy $\overline{S}_B$:
\begin{equation}
S_G = \overline{S}_B - \sum_{a \in A} P(a) \ln P(a) \label{eqgibbsentgen1} .
\end{equation}
The difference between $S_G$ and $\overline{S}_B$ is the second term in (\ref{eqgibbsentgen1}). The interpretation of this term is the following; we see that it has the Gibbs-Shannon form for  entropy of a probability distribution. However, unlike the Gibbs entropy (\ref{eqgibbsentgen}), this is the entropy of a probability distribution $P(a)$ for macroscopic states $a \in A$ of the system. Relation (\ref{eqgibbsentgen1}) is derived with the proviso on the configuration probability $p_{i_a} = p(a)$, $a \in A, i_a = 1, \dots, W(a)$ and puts the mean value of the Boltzmann entropy as well as Gibbs definition of the entropy in the following order $S_G \geq \overline{S}_B$. The exact  equality, in contrast to the one defined  in the thermodynamic limit discussed in following text,  holds in a case when all allowed configurations  belong only to one macroscopic state.

The statistical description relies on the assumption that relative fluctuations of the macroscopic quantities vanish. Macroscopic states are determined by the values of a set of macroscopic quantities. It means that in the thermodynamic limit the probability $P(a)$ of the macroscopic state becomes sharply and overwhelmingly distributed around the most probable macroscopic state characterized by the mean values of macroscopic quantities. Therefore, in the thermodynamic limit, the second term in (\ref{eqgibbsentgen1}) becomes negligible relative to the mean Boltzmann entropy and the Gibbs entropy, which can be taken to be equal. The second term in (\ref{eqgibbsentgen1}) can be called entropy of fluctuations of the macroscopic state, or entropy of fluctuations. 

However, there are exceptions, when the fluctuations of the macroscopic quantities do not vanish in the thermodynamic limit. The simplest example is the ferromagnetic Ising model in the  zero external  magnetic field, where at temperatures below the critical temperature $T < T_c$ almost all spins tend to spontaneously line up, either up or down, with equal probabilities. Therefore, for $T < T_c$ there is a negligible probability of the magnetization ever being equal to its mean value, which is zero because of up-down symmetry of the system. Consequently, for $T < T_c$, the fluctuations of the magnetization do not approach zero in the thermodynamic limit 
 \cite{Garrod}. This is just an example of the behavior near the critical point where the fluctuations of the densities become very large. In such cases, entropy of fluctuations does not become negligible in the thermodynamic limit and, according to (\ref{eqgibbsentgen1}), the mean Boltzmann entropy $\overline{S}_B$ is not close to the Gibbs entropy $S_G$. 

This problem can be partially remedied by introducing a weak symmetry-breaking field. In the context of ferromagnetic Ising model, it is a weak external magnetic field that creates an energy difference between the up-magnetized and down-magnetized state. This modification of the standard statistical approach eliminates the large fluctuations of magnetization, and also the above mentioned discrepancy of predictions compared to physical observations as explained in the reference \cite{Garrod}. An illustrative example is also a water-vapour system in zero gravity  which is characterized by large fluctuations of water position within the hydrophobic container \cite{Garrod}. The measured density at different locations within the container is either the density of water or saturated vapour, in contrast to a uniform average density, intermediate between the water density and vapour density,  predicted by  statistical physics. Evidently standard statistical approach gives a result that does not agree with the experiment. In order to fix this, a very weak but finite gravitational field should be introduced.  However, near a  critical point, to obtain correct predictions standard statistical approach should still be  extended using a different, renormalization group approach \cite{Wilson,Kadanoff}.     

In the next sections we will discuss two types of systems with distinguishable and independent particles: i) the system of $N$ particles with two equally probable single particle states, and ii) the system of $N$ particles with $m$ single particle states with nonuniform probabilities. Using these two examples we will show that for all systems described by a multinomial probability for single particle state occupation numbers, the entropy of fluctuations becomes negligible relative to $\overline{S}_B$ and $S_G$ which become equal in the thermodynamic limit.

\section{A system of $N$ distinguishable particles with two equally probable states} \label{secSNDPWTEPS}

For a system of $N$ distinguishable and independent particles each with two possible states, the multiplicity factor, i.e. the number of configurations - possible realizations of the macroscopic state with $N_1$ and $N_2 $ ($N_1 + N_2 = N$) particles in the two single particle states, is given by    
\begin{eqnarray}
W(N_1, N_2) = \frac{N!}{N_1 ! N_2 !} = \frac{N!}{\left (\frac{N}{2} - n \right )! \left (\frac{N}{2} + n \right )!} = W(n) . \label{eqmulfacbin}
\end{eqnarray}
Here, $n = - \frac{N}{2}, \dots, \frac{N}{2}$ is a deviation of the single particle occupation numbers $N_1$, $N_2$ from the macroscopic state with the maximum multiplicity, which is obtained for $N_1 = N_2 = \frac{N}{2}$. The simplest example is given by the system of $N$ independent classical molecules held in a container which is partitioned only virtually in two fully permeable halves. Another example is the system of $N$ independent distinguishable quantum particles, where each particle can be only in two possible states. Such a system is well approximated by a dilute gas or a crystal lattice where the atoms, each having only two possible values of the $z$ component of magnetic moment $\mu $ and $-\mu $, are well separated and without wave function overlap and particle exchange. For this reason, they can be considered to be independent and distinguishable. A common characteristic of these systems is a possibility to associate a number with each distinguishable particle. A configuration minds about the number of particles in each compartment (or a state) as well as about a number of each particle.

If, for each particle, two possible single particle states have equal probability $\frac{1}{2}$, then the probability for each one of the total of $2^N$ different configurations for a system of $N$ such particles is $\frac{1}{2^N}$. Hence, the probability of the macroscopic state characterized by the deviation $n$ from the most probable macroscopic state is given by      
\begin{eqnarray}
P(n) = \frac{W(n)}{2^N} = \frac{N!}{\left (\frac{N}{2} - n \right )! \left (\frac{N}{2} + n \right )!}\frac{1}{2^N} . \label{eqbindist}
\end{eqnarray}
In probability theory, this is known as the binomial distribution for the number of occurrences $N_1$ and $N_2$ of two mutually exclusive events, obtained in $N_1 + N_2 = N$ repeated independent trials, for the special case when the probabilities of the two events are equal $p_1 = p_2 = \frac{1}{2}$. The binomial distribution can be generalized to the case when there are $m$ mutually exclusive events or outcomes in each of the $N$ repeated independent trials; this generalization is known as the multinomial distribution, elaborated in \ref{multdist} and used throughout of this paper. 

The multiplicity factor (\ref{eqmulfacbin}) for the binomial distribution is also called the binomial coefficient. In the case when there are $m$ mutually exclusive events in each of the $N$ repeated random trials, the multiplicity factor, i.e. the number of possible realizations of such an experiment is given by the generalization, known as the multinomial coefficient (\ref{eqmulfac}) given in \ref{multcoeff}. For the case when the probabilities of all of the total of $m$ mutually exclusive events are equal $p_1 = \cdots = p_m = \frac{1}{m}$, the maximum of the multinomial coefficient (\ref{eqmulfac}) and the maximum of the multinomial distribution (\ref{eqmultdist}) coincide at $N_1 = \cdots = N_m = \frac{N}{m}$. Analogously, for the binomial coefficient and binomial distribution, in case when two mutually exclusive events have equal probabilities $p_1 = p_2 = \frac{1}{2}$, the maxima for both coincide at $N_1 = N_2 = \frac{N}{2}$. The reader can easily check these simple but very important facts. 

For small deviations $n_1= -n$ and $n_2 = n$ of the single particle occupation numbers $N_1 = \frac{N}{2} + n_1$ and $N_2 = \frac{N}{2} + n_2$, the logarithm of the binomial coefficient is appropriately expanded up to the order $n^2$ from the maximum situated at $N_1 = N_2 = \frac{N}{2}$. This is obtained using the analogous expansion (\ref{eqexplogmultcoeffsd}) for the logarithm of the multinomial coefficient derived in \ref{multcoeff}, and putting, for the case at hand given here, $m=2$, $p_1 = p_2 = \frac{1}{2}$ and $n_1= -n$ and $n_2 = n$: 
\begin{equation}
\ln W(n) \approx   (N + 1)\ln 2 - \frac{1}{2}\ln N - \frac{1}{2} \ln (2\pi ) - \frac{N}{2}\left ( \frac{2n}{N}\right )^2 . \label{eqexlogbincoeff}
\end{equation}
For a large number of particles $N >> 1$ and small deviations $\left |\frac{2n}{N} \right | << 1$ this approximation is sufficient. Furthermore, by exponentiating Eq. (\ref{eqexlogbincoeff}) one obtains 
\begin{equation}
W(n) =    2^N \sqrt{\frac{2}{\pi N}} e^{-\frac{2n^2}{N}} . \label{eqexbincoeff}
\end{equation}
Then, using Eq. (\ref{eqbindist}) it is easy to see that fluctuations $n_1= -n$ and $n_2 = n$ of the single particle occupation numbers $N_1 = \frac{N}{2} + n_1$ and $N_2 = \frac{N}{2} + n_2$ are approximately Gaussian distributed around most probable values $N_1 = N_2 = \frac{N}{2}$:  
\begin{equation}
P(n) = \sqrt{\frac{2}{\pi N}}e^{-\frac{2n^2}{N}} .  \label{eqgaussappbindist}
\end{equation}
Furthermore, we can pass on to a continuum approximation and replace the discrete distribution (\ref{eqbindist}) by a continuous Gaussian form (\ref{eqgaussappbindist}), which, as it is written, is properly normalized. The standard deviation, or root-mean-square fluctuation, for the Gaussian distribution (\ref{eqgaussappbindist}) is 
\begin{equation}
\sqrt{\overline{n^2}} = \frac{\sqrt{N}}{2} . \label{eqstatdevbin}
\end{equation}
Interestingly, in the case of a discrete distribution (\ref{eqbindist}), the same result $\sqrt{\overline{n^2}} = \frac{\sqrt{N}}{2}$ for the standard deviation is obtained. One can easily verify this using the expression for the square of standard deviation (\ref{eqsqrtstadevmult}) of the multinomial distribution (\ref{eqmultdist}) in \ref{multdist}, for the case $m=2$, $p_1 = p_2 = \frac{1}{2}$. 

Again, we emphasize that the Gaussian approximation (\ref{eqgaussappbindist}) is valid for small fluctuations $\left |\frac{2n}{N} \right | << 1$ and large number of particles $N >> 1$. However, for $N >> 1$, large fluctuations around the most probable values of occupation numbers $N_1 = N_2 = \frac{N}{2}$ have a very small probability, as is easily seen from  (\ref{eqbindist}) and (\ref{eqgaussappbindist}). This is due to the multiplicity factor $W\left (n = \frac{xN}{2}\right )$, which for large $N$ has a very sharp spike at $x = 0$. The relative width of the spike is given by 
\begin{equation}
\Delta x = \frac{\sqrt{\overline{n^2}}}{\frac{N}{2}} = \frac{1}{\sqrt{N}} , 
\end{equation}
and in this sense it becomes narrower as the number of particles $N$ becomes larger. Consequently for $N >> 1$ this is the reason why $W\left (n = \frac{xN}{2}\right )$ and $P\left (n = \frac{xN}{2}\right )$ are well approximated even for larger fluctuations compared to $\left |\frac{2n}{N} \right | << 1$ by the functions given by (\ref{eqexbincoeff}) and (\ref{eqgaussappbindist}).

A configuration is more (in case of the system of crystal with molecules in nonoverlapping electron magnetic states) or less (in case of classical molecules in the container) counterpart to the microscopic state of the physical system. We exploit the Boltzmann definition of entropy, which for a system of $N$ independent and distinguishable particles with two single particle states, apart from the Boltzmann constant $k$,  is 
\begin{equation}
S_B = \ln W(n) , \label{eqboltentbin}
\end{equation}
where $W(n)$ is the multiplicity (\ref{eqmulfacbin}) of this macroscopic state. Mean value of the Boltzmann entropy (\ref{eqboltentbin}) is then obtained by averaging $\ln W(n)$ over all possible macroscopic states whose probabilities are distributed by $P(n)$: 
\begin{equation}
\overline{S}_B = \sum_{n = -\frac{N}{2}}^{\frac{N}{2}} P(n)\ln W(n) . \label{eqmeanboltentbin}
\end{equation}
Using (\ref{eqexlogbincoeff}) and the normalization property of the probability distribution $P(n)$, from (\ref{eqmeanboltentbin}) one obtains
\begin{equation}
\overline{S}_B \approx  (N + 1)\ln 2 - \frac{1}{2}\ln N - \frac{1}{2} \ln (2\pi ) - \frac{2}{N} \overline{n^2} . \label{eqmeanboltentbin1}
\end{equation}
Introducing (\ref{eqstatdevbin}) in (\ref{eqmeanboltentbin1}) we finally obtain 
\begin{equation}
\overline{S}_B =  (N + 1)\ln 2 - \frac{1}{2}\ln N - \frac{1}{2} \ln (2\pi ) - \frac{1}{2} \approx \ln (2^N) . \label{eqmeanboltentbin2}
\end{equation}
This is the mean value of the Boltzmann entropy for a system of $N$ independent and distinguishable particles with two equally probable single particle states; for a large number of particles $N >> 1$ we can take that it is equal to $N\ln 2$. The value of the Boltzmann entropy for a macroscopic state characterized by occupation numbers $N_1 = \frac{N}{2} - n$ and $N_2 = \frac{N}{2} + n$ is obtained from (\ref{eqexlogbincoeff}) and (\ref{eqboltentbin}): 
\begin{equation}
S_B = (N + 1)\ln 2 - \frac{1}{2}\ln N - \frac{1}{2} \ln (2\pi ) - \frac{2n^2}{N} = \left (S_B\right )_{\mathrm{max}} - \frac{2n^2}{N} . \label{eqboltentbinsmf}
\end{equation}
Equation (\ref{eqboltentbinsmf}) is the expansion of $S_B$ around the maximum value 
\begin{equation}
\left (S_B\right )_{\mathrm{max}} = (N + 1)\ln 2 - \frac{1}{2}\ln N - \frac{1}{2} \ln (2\pi ) \approx  \ln \left (2^N \right ) , \label{eqboltmaxentbin}
\end{equation}
valid for large number of particles $N >> 1$ and small fluctuations $\left |\frac{2n}{N} \right | << 1$.

Knowing the exact configuration for a system of distinguishable independent particles, specifies on the other hand, not only the single particle state occupation numbers, but also the state in which each of the distinguishable particles is found. Accordingly, for a system of $N$ distinguishable particles with two equally probable single particle states, the probabilities of individual configurations of a system are given by
\begin{equation}
\mathcal {P}_i = \frac{1}{2^N} , \label{eqmicroprobbin}
\end{equation}
where configurations are denoted by $i=1, \dots, 2^N$. Then, according to the Gibbs definition of entropy, we have that its value for the configuration probability (\ref{eqmicroprobbin}) is equal to
\begin{eqnarray}
S_G = - \sum_{i=1}^{2^N} \mathcal {P}_i \ln \mathcal{P}_i = \ln (2^N) . \label{eqgibbsentropybin}
\end{eqnarray}

The equations (\ref{eqexbincoeff}), (\ref{eqgaussappbindist}) and (\ref{eqboltentbinsmf}) tell us that, for large $N$, the probability of fluctuations of the Boltzmann entropy $S_B$ is exponentially (Gaussian) distributed around its maximum value $(S_B)_{\mathrm{max}} = \ln (2^N)$, which is also its most probable value. Furthermore, this value is also, according to (\ref {eqgibbsentropybin}), equal to the Gibbs entropy $S_G$:
\begin{equation}
(S_B)_{\mathrm{max}} = \ln (2^N) = S_G . \label{eqentboltmaxentgibbsbin}
\end{equation}
The mean value of the Boltzmann entropy $\overline {S}_B$, according to (\ref{eqmeanboltentbin2}) and (\ref{eqentboltmaxentgibbsbin}), becomes equal to $(S_B)_{\mathrm{max}}$ and $S_G$ in the limit of large number of particles $N >> 1$.

Using (\ref{eqmeanboltentbin2}) and (\ref{eqboltmaxentbin}), one obtains the mean relative fluctuation of the Boltzmann entropy $S_B$ around its maximum and most probable value, 
\begin{equation}
\frac{\overline{S}_B - (S_B)_{\mathrm{max}}}{(S_B)_{\mathrm{max}}} = \frac{ - \frac{1}{2}}{N \ln 2} \propto \frac{1}{N},
\end{equation}
valid in the limit of large number of particles $N >> 1$.

\section{Boltzmann and Gibbs entropy for nonuniform single particle state probabilities $p_1, \dots, p_m$} 
\label{Nonuniform}

Now, consider a case of $N$ distinguishable and independent particles with nonuniform probabilities $p_1, \dots, p_m$ of $m$ single particle states. In analogy to derivation of (\ref{eqbindist}), using (\ref{eqmultdist}), (\ref{eqoccnumdevnoneqprob}) and (\ref{eqmulfacdev}) in the general case of nonuniform probabilities $p_1, \dots, p_m$, we can write for the probability $P(n_1, \dots , n_m)$ of the macroscopic state characterized by the deviations $n_i = N_i - \overline {N_i}$ of occupation numbers from their mean values $ \overline {N_i} = Np_i$, $i = 1, \dots, m$: 
\begin{equation}
P(n_1, \dots , n_m)  =  W(n_1, \dots , n_m) p (n_1, \dots , n_m) . \label{eqmultidistdevnoneqprob}
\end{equation}
Here, probability $P(n_1, \dots , n_m)$ of the macroscopic state is given as a product of the multiplicity $W(n_1, \dots , n_m)$ of the macroscopic state,
 and equal probability $p (n_1, \dots , n_m)$ of configurations corresponding to it. The probability of corresponding configurations is given by
\begin{eqnarray}
p (n_1, \dots , n_m) = \prod_{i=1}^m p_i^{Np_i + n_i } = \exp \left [\sum_{i=1}^m \left (Np_i + n_i \right )\ln p_i\right ] . \label{eqsingleconfigdevnoneqprob}
\end{eqnarray}
For large number of particles $N >> 1$ and small deviations $\left |\frac{n_i}{Np_i} \right | << 1$, using (\ref{eqexplogmultcoeffsd}) we easily obtain
\begin{eqnarray}
W(n_1, \dots , n_m)  =  \frac{N!}{\prod_{i=1}^m (Np_i + n_i) !} \cr\nonumber\\
 \approx  \sqrt{\frac{N}{(2\pi)^{m-1} \prod_{i=1}^m Np_i}}\exp \left [ - \sum_{i=1}^m \left (Np_i + n_i \right )\ln p_i -\frac{1}{2} \sum_{i=1}^m \frac{n_i^2}{N p_i}\right ] . \label{eqexpmultcoeffnoneqgauss}
\end{eqnarray}
Then, using (\ref{eqmultidistdevnoneqprob}), (\ref{eqsingleconfigdevnoneqprob}) and (\ref{eqexpmultcoeffnoneqgauss}) we get the macrostate probability in the Gaussian approximation,
\begin{equation}
P(n_1, \dots , n_m) = \sqrt{\frac{N}{(2\pi)^{m-1} \prod_{i=1}^m Np_i}} \exp \left ( -\frac{1}{2} \sum_{i=1}^m \frac{n_i^2}{N p_i}\right ) . \label{eqmultdistnoneqprobgauss}
\end{equation}
Eqs. (\ref{eqmultdistnoneqprobgauss}) and (\ref{eqdevoccupnumnon}) tell us that the macroscopic state characterized by the occupation numbers equal to their mean values $ \overline {N_i} = Np_i$, $i = 1, \dots, m$, is the most probable macroscopic state. On the other hand, the macroscopic state with occupation numbers $N_{i} = N/m$ for all $i = 1, \dots, m$ has the greatest multiplicity (\ref{eqmulfac}). In the case when single particle state probabilities are equal $p_1, \dots, p_m = 1/m$, this is also the most probable state, as is obvious from (\ref{eqmultdist}) and (\ref{eqmulfac}). However, for nonuniform single state probabilities $p_1, \dots, p_m$, the macroscopic state with the greatest multiplicity and the most probable macroscopic state do not coincide, in contrast to the case of uniform single state probabilities. 

As physically very interesting example, we choose Gibbs canonical distribution. In case of a system of $N$ distinguishable and independent particles with $m$ single particle states, probabilities of configurations that correspond to the macroscopic state with occupation numbers $N_1, \dots, N_m$ are equal  
\begin{equation}
p (N_1, \dots , N_m) = \frac{\prod_{i=1}^m e^{-\frac{N_i E_i}{kT}}}{Z^N} . \label{eqgibbscannconfig}
\end{equation}
Here, $Z$ is a single particle partition function
\begin{equation}
Z = \sum_{i=1}^m e^{-\frac{E_i}{kT}} ,
\end{equation}
and $E_i$ is the energy of the $i$-th single particle state. The partition function of the system of $N$ particles is then the $N$-th power of the single particle partition function $Z$. 

We note here that the configuration probability (\ref{eqgibbscannconfig}) comes from (\ref{eqsingleconfigdevnoneqprob}), too. It follows simply by rewriting (\ref{eqsingleconfigdevnoneqprob}) in terms of occupation numbers introduced in (\ref{eqoccnumdevnoneqprob}) and using the Gibbs single particle state probabilities
\begin{equation}
p_i = \frac{e^{-\frac{E_i}{kT}}}{Z} ,
\end{equation}
for $i = 1, \dots, m$. Macrostate probability (\ref{eqmultidistdevnoneqprob}), written in the form which uses the occupation numbers $N_1, \dots, N_m$ as the variables, reads
\begin{eqnarray}
P(N_1, \dots , N_m) = W(N_1, \dots , N_m) p (N_1, \dots , N_m) . \label{eqmacroproboccupnum}
\end{eqnarray}
Then using (\ref{eqmacroproboccupnum}) and (\ref{eqmulfac}) it is easy to obtain, for the example of Gibbs configuration probability (\ref{eqgibbscannconfig}), the probability of the macroscopic state characterized by single particle state occupation numbers $N_1, \dots, N_m$.

It is convenient, for nonuniform probabilities $p_1, \dots, p_m$ of single particle states to write the occupation numbers in the form used in (\ref{eqoccnumdevnoneqprob}). The mean value of Boltzmann entropy is then obtained by averaging $S_B = \ln W(n_1, \dots , n_m)$ over the macrostate probability $P(n_1, \dots , n_m)$ given by (\ref{eqmultidistdevnoneqprob}). We use the small fluctuation approximation of $\ln W(n_1, \dots , n_m)$ given by the expansion (\ref{eqexplogmultcoeffsd}) around the most probable state, characterized by occupation numbers equal to their mean values $ \overline {N_i} = Np_i$, $i = 1, \dots, m$.  Using (\ref{eqmultdistfluct}), we then obtain
\begin{eqnarray}
\overline{S}_B & = & - N \sum_{i=1}^m p_i \ln p_i +\frac{1}{2}\ln N - \frac{1}{2}\sum_{i=1}^m\ln (Np_i) \nonumber\\ 
& & - \frac{m-1}{2}\ln(2\pi) -\frac{1}{2} \sum_{i=1}^m \frac{\overline {n_i^2}}{N p_i} . \label{eqmeanboltentmultinoneq1}
\end{eqnarray}
For the squares of standard deviations (variances) $\overline {n_i^2}$ we use (\ref{eqsqrtstadevmult}) and finally, using the normalization property of single particle state probabilities, equation (\ref{eqmeanboltentmultinoneq1}) becomes
\begin{eqnarray}
\overline{S}_B & = &  - N \sum_{i=1}^m p_i \ln p_i +\frac{1}{2}\ln N - \frac{1}{2}\sum_{i=1}^m\ln (Np_i) - \frac{m-1}{2}
[\ln(2\pi) + 1] \nonumber\\
& \approx & - N \sum_{i=1}^m p_i \ln p_i. \label{eqmeanboltentmultinoneq2}
\end{eqnarray}
Here, for large $N >> 1$, the mean value of Boltzmann entropy of the system of $N$ distinguishable independent particles with $m$ single particle states with probabilities $p_1, \dots, p_m$ is equal to the last expression in (\ref{eqmeanboltentmultinoneq2}).

Now, we show briefly that the last expression in (\ref{eqmeanboltentmultinoneq2}) is equal to the Gibbs entropy of this system. Probability of an individual configuration for the system of $N$ distinguishable independent particles with single particle state probabilities $p_1, \dots, p_m$ is given by the corresponding product of $N$ of these probabilities. The product is determined by a state of each individual particle in this configuration. This is written as
\begin{equation}
\underbrace{\mathcal{P}_{i, j, \dots}}_{N \ indices} =  \underbrace {p_i p_j \cdots }_{N \ times} , \label{eqmicroprobmultinoneq}
\end{equation}
where each of $N$ indices denotes the state of an individual particle,  and all indices go from $i, j ,\dots = 1, \dots, m$, thus denoting $m^N$ different configurations of a system. According to the Gibbs definition of entropy, for configuration probability (\ref{eqmicroprobmultinoneq}) it is equal to
\begin{equation}
S_G = - \sum_{\underbrace{i, j, \dots}_{N \ indices} =1}^m \mathcal{P}_{i, j, \dots} \ln \mathcal{P}_{i, j, \dots} = - N \sum_{i=1}^m p_i \ln p_i  . \label{eqgibbsentropymultinoneq}
\end{equation}
By comparing (\ref{eqmeanboltentmultinoneq2}) and (\ref{eqgibbsentropymultinoneq}) we see that, 
\begin{equation}
S_G = \overline{S}_B - \frac{1}{2}\ln N + \frac{1}{2}\sum_{i=1}^m\ln (Np_i) + \frac{m-1}{2}
[\ln(2\pi) + 1] , \label{eqgibbsentmeanboltentgeneq}
\end{equation}
and furthermore, that for large number of particles $N >> 1$, the mean value of the Boltzmann entropy becomes equal to the Gibbs entropy, 
\begin{equation}
\overline{S}_B \approx  S_G . \label{eqmeanboltentapproxgibbsent}
\end{equation}
Equations (\ref{eqgibbsentmeanboltentgeneq}) and (\ref{eqmeanboltentapproxgibbsent}) are general equations valid for systems of $N$ distinguishable independent particles with $m$ single particle states, and therefore they are also valid for the special case discussed in the previous section. Clearly, they represent a consequence of the negligibility of the entropy of fluctuations in equation (\ref{eqgibbsentgen1}) in the thermodynamic limit.

However, as it has been pointed out in this section, for nonuniform single particle state probabilities, the macroscopic state with maximum multiplicity (and therefore with the maximum Boltzmann entropy), and the most probable macroscopic state do not coincide. From (\ref{eqmultdistnoneqprobgauss}) and (\ref{eqexplogmultcoeffsd}), we see that in the small fluctuation expansion of $S_B = \ln W(n_1, \dots , n_m)$ around the most probable macroscopic state characterized by $n_1 = \cdots = n_m = 0$, the Boltzmann entropy is equal to 
\begin{eqnarray}
S_B & = & - N \sum_{i=1}^m p_i \ln p_i +\frac{1}{2}\ln N - \frac{1}{2}\sum_{i=1}^m\ln (Np_i) - \frac{m-1}{2}\ln(2\pi) \nonumber\\
& & - \sum_{i=1}^m n_i \ln  p_i -\frac{1}{2} \sum_{i=1}^m \frac{n_i^2}{N p_i} . \label{eqboltentmultinoneqprob}
\end{eqnarray}
The value of the Boltzmann entropy $S_B$ fluctuates, with probability approximated by (\ref{eqmultdistnoneqprobgauss}), around its most probable value for $n_1 = \cdots = n_m = 0$, which is equal to the first line in (\ref{eqboltentmultinoneqprob}). The most probable value of $S_B$ is associated to the most probable macroscopic state, where the system described by the distribution (\ref{eqmultidistdevnoneqprob}) is expected to spend much more time than in any other macroscopic state. Of course, this expectation is in some extent analogous to the ergodic hypothesis, that the time averages of physical quantities for a system described by microcanonical ensemble coincide with their statistical averages, and therefore that they are very close to their statistically most probable values. However, concepts of ergodic theory are a topic for themselves and too complex to be discussed in this paper.  This is particularly true for systems not described by a microcanonical distribution, like the ones described by equilibrium Gibbs canonical distribution. More importantly, we have shown that for a system of $N$ distinguishable independent particles with general single particle state probabilities, the most probable value of the Boltzmann entropy $S_B$ becomes equal to the Gibbs entropy $S_G$,
\begin{equation}
\left ( S_B \right)_{\mathrm{most \ prob.}} = S_G = - N \sum_{i=1}^m p_i \ln p_i . \label{eqboltentmostprobeqgibbs}
\end{equation}
in the limit of large number of particles $N >> 1$. In the same limit, according to (\ref{eqmeanboltentmultinoneq2}) and (\ref{eqboltentmostprobeqgibbs}), the mean value $\overline{S}_B$ of Boltzmann entropy becomes equal to them. The mean relative fluctuation of the Boltzmann entropy around its most probable value is 
\begin{equation} 
\frac{\overline{S}_B - \left ( S_B \right)_{\mathrm{most \ prob.}}}{\left ( S_B \right)_{\mathrm{most \ prob.}}} = \frac{ - \frac{m-1}{2}}{- N \sum_{i=1}^m p_i \ln p_i} \propto  - \frac{1}{N} ,
\end{equation}
and it vanishes for $N >> 1$.

\section{Conclusion}

It is widely accepted to extrapolate the Clausius definition of an equilibrium state of an isolated, or more precisely thermally isolated, system as the state with maximum entropy. Strictly speaking, from the statistical point of view, a system in equilibrium does not find itself in the state of maximum but rather mean entropy. Averaging the Boltzmann entropy with respect to probabilities of macroscopic states, we come to the equation (\ref{eqgibbsentgen1}) which is the central result of this work. It tells us that mean Boltzmann entropy is equal to the Gibbs entropy minus a quantity that includes fluctuations of macroscopic physical quantities. This quantity has the form of Gibbs entropy and we have named it entropy of fluctuations. The equality between mean Boltzmann and Gibbs entropy depends on the ratio between fluctuation entropy and mean Boltzmann entropy. If this ratio converges to zero in thermodynamic limit the mean Boltzmann entropy becomes equal to the Gibbs entropy.
 
In this work we have considered a system of $N$ distinguishable and independent particles
with $m$ single particle states. It is shown that the ratio between the fluctuation entropy and the most probable value of Boltzmann entropy vanishes as $1/N$ in the thermodynamic limit. In this way there is no practical difference between mean Boltzmann and Gibbs entropy. 
 
From the pedagogical reasons, as a special case, the system of particles with two equally probable single particle states was considered first.  This case has a common characteristic with the microcanonical ensemble and uniform ensemble that all configurations available to the system are equally probable. It is found that mean Boltzmann entropy converges to maximum entropy in the thermodynamic limit, in accordance with Clausisus inequality. 
 
In the case of nonuniform single particle state probabilities, mean Boltzmann entropy is no longer close to its maximum value, as it is expected, but it converges to Gibbs entropy. 
 
It comes out that the ratio between fluctuation entropy and mean Boltzmann entropy is a criterion of validity of the standard statistical physics approach. If this ratio does not vanish in the thermodynamic limit, standard statistical approach  should be modified, or extended with other methods like renormalization group.

 It is pointed out that the ferromagnetic Ising model in the zero external magnetic field is an example of system with large fluctuation entropy for temperatures close and below the critical one. It is self-evident that  standard statistical approach can give nonphysical values of mean magnetization in this case.
 
 Additional cases are materials close to critical point in the $(p,V)$ diagram. The critical opalescence is the result of large fluctuation entropy. One expects that critical phenomena in general are characterized by states of large fluctuation entropy.

\appendix
\section{Multinomial distribution} \label{multdist}

Here we use definitions from reference \cite{MD}.

Multinomial distribution  \cite{MD} is a joint distribution of random variables that is defined for all nonnegative integers $N_1, \dots, N_m$ satisfying the condition $\sum_{i=1}^{m}N_i = N$, $N_i = 0, \dots, N, i = 1, \dots, m$, by the formula
\begin{equation}
P(N_1, \dots , N_m) = \frac{N!}{\prod_{i=1}^m N_i !}\prod_{i=1}^m p_i^{N_i} \label{eqmultdist} .
\end{equation}
Parameters of the distribution are $N$ and $p_1, \dots, p_m$, and satisfy the condition $\sum_{i=1}^m p_i = 1$, $p_i \geq 0, i = 1, \dots, m$. In probability theory, (\ref{eqmultdist}) is the probability for the number of occurrences $N_1, \dots, N_m$ of $m$ mutually exclusive events, obtained in $\sum_{i=1}^{m}N_i = N$ repeated independent trials, where the probabilities of the events in each trial are $p_1, \dots, p_m$. Multinomial distribution (\ref{eqmultdist}) is a natural generalization of the binomial distribution, which is its special case for $m = 2$.

The mathematical expectations of $N_i$ are, for $i = 1, \dots, m$, given by
\begin{equation}
\overline {N_i} = \sum_{{N_1, \dots , N_m = 0, \atop \sum_{i=1}^{m}N_i = N}}^{N} N_i P(N_1, \dots , N_m) = Np_i . \label{eqavgoccupnumnon}
\end{equation}
We define also, for $i = 1, \dots, m$, the deviations (or fluctuations) of $N_i$ from their mean values 
\begin{equation}
n_i = N_i - \overline {N_i} . \label{eqdevoccupnumnon}
\end{equation}
Obviously, the mathematical expectations of  $n_i$ are, for $i = 1, \dots, m$,
\begin{equation}
\overline {n_i} = 0 \label{eqmultdistfluct} .
\end{equation}
We also define the mathematical expectations of $n_i^2$, or equivalently, the variances of $N_i$. For  $i = 1, \dots, m$, they are equal to 
\begin{equation}
\overline {n_i^2} = \sum_{{N_1, \dots , N_m = 0, \atop \sum_{i=1}^{m}N_i = N}}^{N} \left (N_i - \overline {N_i} \right )^2 P(N_1, \dots , N_m) = \overline {N_i^2} - \overline {N_i}^2 .
\end{equation}
Specifically, when they are taken with respect to the multinomial distribution (\ref{eqmultdist}), one obtains 
\begin{equation}
\overline {n_i^2}  = Np_i \left (1 - p_i \right ) . \label{eqsqrtstadevmult}
\end{equation}

\section{Multinomial coefficient} \label{multcoeff}

In this appendix we use definitions from references \cite{MD} and \cite{MC}.

In combinatorics the multinomial coefficient expresses \cite{MC} the number of ways of putting $N$ different elements in $m$ different boxes in which box $i$ contains $N_i$ elements, $i = 1, \dots, m$, without taking the order in which we put the elements into boxes into account. It is given by
\begin{equation}
W(N_1, \dots , N_m) = \frac{N!}{\prod_{i=1}^m N_i !} \label{eqmulfac}
\end{equation}
Now we assume that putting individual elements into boxes are independent events, and that for each element the probabilities for putting it into $m$ different boxes are $p_1, \dots, p_m$, $\sum_{i=1}^m p_i = 1$. Then the probabilities for the number of elements $N_1, \dots, N_m$, $\sum_{i=1}^{m}N_i = N$, in $m$ boxes are distributed according to the multinomial distribution (\ref{eqmultdist}). 

Introducing the deviations (\ref{eqdevoccupnumnon}) of $N_1, \dots, N_m$ from the mean values (\ref{eqavgoccupnumnon}) taken with respect to the multinomial distribution (\ref{eqmultdist}), we can write $N_i$, $i = 1, \dots, m$, as
\begin{equation}
N_i = Np_i + n_i . \label{eqoccnumdevnoneqprob}
\end{equation}
Introducing (\ref{eqoccnumdevnoneqprob}) in the multinomial coefficient (\ref{eqmulfac}) we obtain
\begin{equation}
W(n_1, \dots , n_m) = \frac{N!}{\prod_{i=1}^m \left (Np_i + n_i \right )!} . \label{eqmulfacdev}
\end{equation}
The Stirling formula for $x!$ is valid for large $x$, 
\begin{equation}
\ln \left (x! \right ) \approx  x\ln x - x + \frac{1}{2}\ln x + \frac{1}{2} \ln (2\pi) . \label{eqstirlform}
\end{equation}
For large $N$, such that $\left |\frac{n_i}{Np_i} \right | << 1$ also, the multinomial coefficient (\ref{eqmulfacdev}) can be calculated using (\ref{eqstirlform}). One then obtains
\begin{eqnarray}
\ln W(n_1, \dots , n_m) & = & \ln \left ( N ! \right) - \sum_{i=1}^m \ln\left [\left (Np_i + n_i \right ) ! \right ] \nonumber\\
& \approx & N \ln N - \sum_{i=1}^m \left (Np_i + n_i \right ) \ln \left (Np_i + n_i \right )  \nonumber\\
& & + \frac{1}{2} \ln N - \frac{1}{2}\sum_{i=1}^m \ln \left (Np_i + n_i \right ) - \frac{m-1}{2}\ln(2\pi ). \label{eqmulfacstirlform}
\end{eqnarray}
We can write (\ref{eqmulfacstirlform}) in the form
\begin{eqnarray}
\ln W(n_1, \dots , n_m)  =  \left (N + \frac{1}{2}\right )\ln N - \frac{m-1}{2}\ln(2\pi) \cr\nonumber\\
 - \sum_{i=1}^m N p_i \left (1 + \frac{n_i + \frac{1}{2}}{Np_i} \right ) \left [\ln \left ( N p_i \right )  + \ln \left (1 + \frac{n_i}{Np_i} \right ) \right ] . \label{eqmulfacstirlform1}
\end{eqnarray}
The expansion of $\ln(1 + x)$ around $x=0$ up to the order $x^2$ is valid approximately for $|x| << 1$ and given by 
\begin{equation}
\ln (1 + x) = x - \frac{1}{2}{x^2} + \dots \ . \label{eqtaylorlog1px}   
\end{equation}
Using (\ref{eqtaylorlog1px}) we obtain the expansion of (\ref{eqmulfacstirlform1}) up to the order $n_i^2$,
\begin{eqnarray}
\ln W(n_1, \dots , n_m)   \approx  - N \sum_{i=1}^m p_i \ln p_i +\frac{1}{2}\ln N - \frac{1}{2}\sum_{i=1}^m\ln (Np_i)  \cr\nonumber\\
- \frac{m-1}{2}\ln(2\pi) - \sum_{i=1}^m \left (\ln  p_i + \frac{1}{2Np_i}\right )n_i -\frac{1}{2} \sum_{i=1}^m \left [\frac{1}{N p_i} - \frac{1}{2(Np_i)^2}\right ]n_i^2 . \label{eqexplogmultcoeff}
\end{eqnarray}
Since $N >>1$ and $\left |\frac{n_i}{Np_i} \right | << 1$, $i = 1, \dots, m $, from (\ref{eqexplogmultcoeff}) we obtain 
\begin{eqnarray}
\ln W(n_1, \dots , n_m) & \approx & - N \sum_{i=1}^m p_i \ln p_i +\frac{1}{2}\ln N - \frac{1}{2}\sum_{i=1}^m\ln (Np_i) \nonumber\\
& & - \frac{m-1}{2}\ln(2\pi) - \sum_{i=1}^m n_i \ln  p_i -\frac{1}{2} \sum_{i=1}^m \frac{n_i^2}{N p_i}  . \label{eqexplogmultcoeffsd}
\end{eqnarray}

\end{document}